\documentclass[10pt,final,journal,twocolumn]{IEEEtran}%
\usepackage{amsmath}
\usepackage{graphicx}
\usepackage{latexsym}
\usepackage{amssymb}
\usepackage{cite}
\usepackage{color}
\usepackage{multirow}
\usepackage{algorithmic,algorithm}
\usepackage{footnote}
\usepackage{flushend}

\begin{document}

\title{Energy Harvesting Enabled MIMO Relaying Through Power Splitting}
\author{Jialing Liao, Muhammad R. A. Khandaker,~\IEEEmembership{Member,~IEEE}, and Kai-Kit Wong,~\IEEEmembership{Fellow,~IEEE}\thanks{The authors are with the Department of Electronic and Electrical Engineering, University College London, UK (e-mail: $\rm jialing.liao.14@ucl.ac.uk$, $\rm m.khandaker@ucl.ac.uk$, $\rm kai$-$\rm kit.wong@ucl.ac.uk$).}\thanks{This work is supported by EPSRC under grant EP/K015893/1.}}
\pagestyle{headings}
\maketitle \thispagestyle{empty}
\begin{abstract}
This paper considers a multiple-input multiple-output (MIMO) relay system with an energy harvesting relay node. All nodes are equipped with multiple antennas, and the relay node depends on the harvested energy from the received signal to support information forwarding. In particular, the relay node deploys power splitting based energy harvesting scheme. The capacity maximization problem subject to power constraints at both the source and relay nodes is considered for both fixed source covariance matrix and optimal source covariance matrix cases. Instead of using existing software solvers, iterative approaches using dual decomposition technique are developed based on the structures of the optimal relay precoding and source covariance matrices. Simulation results demonstrate the performance gain of the joint optimization against the fixed source covariance matrix case.
\end{abstract}


\section{Introduction}
Cooperative communication based on relay has been seen as one of the promising techniques since 1970s \cite{swipt_1} to improve network coverage as well as throughput. Since then, considerable work has been done to explore cooperation strategies from various perspectives. To take the advantages of the multiple-input multiple-output (MIMO) technique, e.g. improving spectrum utilization and link reliability, MIMO relay networks were introduced in \cite{swipt_2, swipt_3, swipt_4, ruhul12, ruhul13, ruhul14, ruhul15, ruhul16, ruhul_yue} where the capacity maximization problem under fixed source and relay transmit power thresholds was investigated. In \cite{swipt_2} and \cite{swipt_3}, relay only and joint source and relay optimization schemes were considered with fixed source covariance matrix and arbitrary source covariance matrix, respectively. Then in \cite{swipt_4}, joint source and relay design was investigated for MIMO-OFDM relay networks.

With green communication becoming an important tendency of next generation wireless communication, recently, researchers have started paying attention to the combination of energy harvesting technique and cooperative communication due to the limited battery storage of the relay nodes \cite{ruhul_kk, ruhuul2, RF2, ruhuul}. In \cite{swipt_5}, the outage probability and the ergodic capacity were analyzed for one-way relaying system with energy harvesting while \cite{swipt_6} focused on the power allocation strategies for multiple source-destination pair cooperative relay networks. In \cite{swipt_7}, energy harvesting was introduced to the cooperative networks with spatially random relays and the outage and diversity performance were investigated using stochastic geometry. In \cite{swipt_8}, the distributed power splitting (PS) based simultaneous wireless information and power transfer (SWIPT) was studied for interference relay channels using game theory. These works only considered SWIPT in single antenna and single-carrier relay networks.

Later in \cite{swipt_9}, SWIPT was considered for a multi-antenna relay network with single antenna source and destination nodes. Aimed to minimize the transmit power at the relay subject to the signal-to-inference-plus-noise ratio (SINR) and energy harvesting constraints with imperfect channel state information (CSI), joint optimization of beamforming and power splitting (PS) ratio was considered in \cite{swipt_9} based on semidefinite programming problem (SDP). In \cite{swipt_10}, the authors focused on the optimal precoding for SWIPT in a two-hop decode-and-forward (DF) MIMO relay network with energy harvesting at the destination. A suboptimal and an iterative approaches guided by semidefinite relaxation (SDR) were developed in \cite{swipt_11} for a half-duplex two-way AF MIMO relay network with a power-splitting based energy harvesting source node aiming to minimize the total mean-squared error (MSE). Note that the works in \cite{swipt_10, swipt_11} provided important results for SWIPT in MIMO relay networks depending on either SDR and existing solvers or iterative methods, but they failed to provide the close-form solutions as well as the structures of the source covariance and relay beamforming matrices. Also, the joint PS ratio and precoding matrices design is not well investigated for generic MIMO relay systems \cite{swipt_10, swipt_11}.

In this paper, power-splitting based energy harvesting is considered for MIMO relay networks. In order to derive the maximum capacity with power constraints at the source and relay nodes, we first consider the case of fixed source covariance matrix to optimize relay beamforming and power splitting ratio at the relay node. Then we jointly design the source covariance matrix, relay beamforming matrix and PS ratio. Instead of using SDR and software solvers, we provide the structures of the optimal source covariance and relay precoding matrix based on which iterative approaches are employed to derive the near-optimal results. Finally, numerical simulations are carried out to investigate the performance of the proposed schemes.

The rest of this paper is organized as follows. In Section~\ref{sec_sys}, the system model of a MIMO relay network with power splitting based energy harvesting relay node is introduced. The fixed source covariance matrix case is elaborated in Section~\ref{sec_algo_relay} while the joint transmit and relay precoding matrices along with power splitting ratio design algorithm is developed in Section~\ref{sec_algo_joint}. Section~\ref{sec_sim} shows the simulation results which justify the significance of the proposed algorithms under various scenarios. Conclusions are drawn in Section~\ref{sec_con}.

\section{System Model}\label{sec_sys}
\begin{figure}
\centering
\includegraphics*[width=8cm]{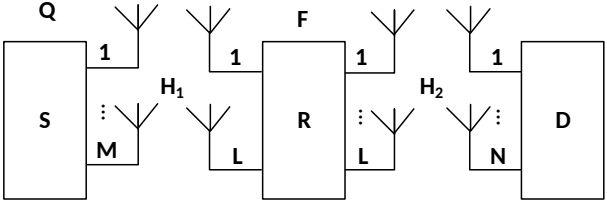}
\caption{System model of a dual-hop MIMO relay network}
\label{sysmod}
\end{figure}

As shown in Fig.~\ref{sysmod}, a two-hop MIMO relay network is considered with power splitting based energy harvesting at the relay node where all the nodes are equipped with multiple antennas. The numbers of antennas for the source, relay and destination nodes are $M,~L$, and $N$, respectively, with the number of transmit data streams $D$ satisfying $~D \le \min(M, L, N)$.

Assuming non-regenerative and half-duplex relaying, the signal transmission can be divided into two phases, the source transmission phase and the relay forwarding phase. In the first phase, information and energy are simultaneously transmitted from the source to the relay while in the second phase the relay forwards the received signals to the destination using the harvested energy from the source. We assume that the source has fixed energy supply and the relay helps forward information to the destination node using the energy harvested from the source in the first phase.

\begin{figure} 
\centering
\includegraphics*[width=8cm]{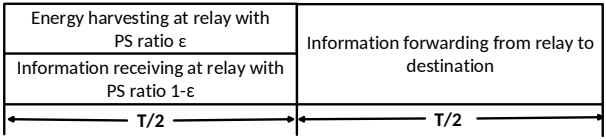}
\caption{The framework of the proposed PSR}
\label{ps}
\end{figure}

The frame of PS based MIMO relay networks is shown in Fig.~\ref{ps}. Let $T$ be the block length and split the time equally between the two phases. $\varepsilon$ is defined as the PS ratio. As can be observed, in the first phase,  $\varepsilon$ of the received signal is used for energy harvesting while the rest is used for information forwarding from the relay to the destination. In the second phase, the information received at relay is forwarded to the destination using the energy harvested in the first phase.

To better clarify the expressions, we summarize some commonly used symbols in Table I.

\begin{table}
  \caption{Symbol Notations}
  \centering 
\begin{tabular}{c l} 
\hline\hline 
Notation & Representation \\ [0.5ex]
\hline 
P & the available transmit power at source   \\ 
${\bf H}_1$ & the channel matrix between the source and relay  \\
${\bf H}_2$ & the channel matrix between the relay and destination  \\
${\bf s}$ & the source symbol vector    \\
${\bf Q}$ & source covariance matrix   \\
${\bf F}$ & relay beamforming matrix   \\
${\bf n}_1$ & the AWGN at relay with a variance of ${\sigma_1}^2$  \\
${\bf n}_2$ & the AWGN at destination with a variance of ${\sigma_2}^2$ \\
$\varepsilon$ & power-splitting ratio \\
\hline
\end{tabular}
\label{table1} 
\end{table}

\section{Optimization with Fixed Source Covariance Matrix}\label{sec_algo_relay}

In the uniform source pre-coding case, i.e. ${\bf Q}=$$\mathbb{E}{({\bf s}{\bf s}^H)}$$=\frac {P}{D}{\bf I}$, the harvested power at relay in the first phase can then be written as
\begin{equation}
{\rm tr}{({\bf y}_e {{\bf y}_e}^H)}={\sigma_1^2}{\rm tr}{(\varepsilon \rho_1 {\bf H}_1{\bf H}_1^H)},
\end{equation}
where we let the received signal for energy harvesting ${\bf y}_e = \sqrt{\varepsilon}{\bf H}_1 s$ and the signal-to-noise ratio (SNR) at the relay $\rho_1\triangleq\frac {P}{D \sigma_1^2}$. Here we assume that the antenna noise is relatively smaller and can therefore be ignored. The received signal at relay for information processing is given by
\begin{equation}
{\bf y}_r=\sqrt{1-\varepsilon}{\bf H}_1 s+{\bf n}_1.
\end{equation}
Then in the second phase, the receive signal at destination is
\begin{equation}
{\bf y}_d = \sqrt{1-\varepsilon}{\bf H}_2 {\bf F} {\bf H}_1 {\bf s}+{\bf H}_2 {\bf F} {\bf n}_1+{\bf n}_2.
\end{equation}
In this case, the achievable rate can be written as
\begin{align}\label{cap}
C =\frac {1}{2}\log_2 {\rm det} \left({\bf I}_D+(1-\varepsilon)\rho_1 {\bf H}_1{\bf H}_1^H \notag \right. \\
\left. -(1-\varepsilon)\rho_1 {\bf H}_1{\bf H}_1^H{\bf W}^{-1} \right),
\end{align}
where ${\bf W}={\bf I}_D+{\bf G}^H{\bf H}_2^H{\bf H}_2 {\bf G},~\text{and}~{\bf G}=\frac {\sigma_1}{\sigma_2}{\bf F}$.

Then we consider the power constraint at relay which is given by
\begin{align}
{\rm tr}{({\bf G}({\bf I}_D+(1-\varepsilon)\rho_1 {\bf H}_1{\bf H}_1^H){\bf G}^H)} \le \eta {\rm tr}{(\varepsilon\rho_2 {\bf H}_1{\bf H}_1^H)},
\end{align}
where $\rho_2\triangleq\frac{\text{P}}{D{\sigma_2}^2}$. $0 \le \eta \le 1 $ denotes the energy convention efficiency. Consequently, the optimization problem of interest is
\begin{subequations}\label{maxRob1}
\begin{align}
\max_{{\bf G},\varepsilon} ~~~C~~~~~~\mbox{s.t.}~~~~~~~~~\\
{\rm tr}{({\bf G}({\bf I}_D+(1-\varepsilon)\rho_1 {\bf H}_1{\bf H}_1^H){\bf G}^H)} \notag\\
\le \eta {\rm tr}{(\varepsilon\rho_2 {\bf H}_1{\bf H}_1^H)}.
\end{align}
\end{subequations}

Clearly, the objective is neither convex nor concave. Hence, the optimization problem can not been solved directly. We consider updating ${\bf G}$ and $\varepsilon$ alternatingly. Let us now define $\hat{\rho}_1=(1-\varepsilon)\rho_1, \hat{\rho}_2=\varepsilon \rho_2$ and fix $ \varepsilon$. The problem then becomes similar to the one in \cite{swipt_2}. Consequently, we consider the singular value decompositions (SVDs) of the channel matrices shown below
\begin{align}
&{\bf H}_1={\bf U}_1 {\bf \Sigma}_1 {\bf V}_1^H, \label{channel} \\
&{\bf H}_2={\bf U}_2 {\bf \Sigma}_2 {\bf V}_2^H. \label{channel1}
\end{align}
where ${\bf \Sigma}_1, {\bf \Sigma}_2$ are diagonal matrices while others are unitary. It can be shown, as in  \cite{swipt_2}, that the optimal relay matrix has structure ${\bf F}={\bf V}_2 {\bf \Lambda }_F {\bf U}_1^H$ where ${\bf \Lambda }_F$ denotes a diagonal matrix. Let ${\bf G}={\bf V}_2 {\bf X}^\frac {1}{2} ({\bf I}+(1-\varepsilon)\rho_1 {\bf \Lambda}_1)^{-\frac {1}{2}} {\bf U}_1^H$ where ${\bf X}$ is a diagonal matrix with ${\bf X}={\rm diag}({x_1,x_2,...,x_D})$. In addition, we let ${\bf \Lambda}_1={\bf \Sigma}_1^2$, ${\bf \Lambda}_2={\bf \Sigma}_2^2$ be diagonal matrices with the vectors $ \boldsymbol{\alpha}=[\alpha_1,...\alpha_D]$ and $\boldsymbol{\beta}=[\beta_1,...\beta_D]$ as the diagonal, respectively. Problem (\ref{maxRob1}) then becomes a scalar optimization problem:
\begin{subequations}\label{maxRob_2_3}
\begin{align}
&~~~~\max_{0 \le \varepsilon \le 1, \{x_k\}}~~f(\{x_k\},\varepsilon)  \label{R_2_3_o}\\
&{\rm s.t.}~~g(\{x_k\},\varepsilon)\triangleq\eta \sum_{k=1}^D \varepsilon\rho_2\alpha_k-\sum_{k=1}^D x_k \ge 0, \label{R_2_2_1} \\
&~~~~~~~x_k \ge 0,~~\forall k
\end{align}
\end{subequations}
where
\begin{align}
f(\{x_k\},\varepsilon)\triangleq&\frac {1}{2}\left[\sum_{k=1}^D \log_2 (1+(1-\varepsilon)\rho_1\alpha_k) \right. \notag \\
&\left. +\sum_{k=1}^D \log_2 \left(\frac {1+\beta_k x_k}{1+(1-\varepsilon)\rho_1\alpha_k+\beta_k x_k}\right)\right].
\end{align}
Considering the Lagrangian, the dual problem can be expressed as
\begin{subequations}\label{maxRob3}
\begin{align}
\max_{\{x_k\},\varepsilon,\atop \nu,\{\lambda_k\}} {\cal L}\triangleq f(\{x_k\},\varepsilon)+\nu g(\{x_k\},\varepsilon)+\sum_{k=1}^D \lambda_k x_k \\
{\rm s.t.}~~0 \le \varepsilon \le 1, \nu \ge 0, x_k \ge 0, \lambda_k \ge 0, \forall k.
\end{align}
\end{subequations}
					
Based on the Karush-Kuhn-Tucker (KKT) conditions, we have
\begin{subequations}\label{minRob6}
\begin{align}
\nu g(\{x_k\},\varepsilon) =0,&\label{R6_o}\\
\lambda_k x_k=0,&\forall k, \label{R6_1}\\
{\nabla_{x_k}}{\cal L}=0,&\forall k, \label{R6_2}\\
\nabla_\varepsilon{\cal L}=0. &\label{R6_3}
\end{align}
\end{subequations}
Using (\ref {R6_2}), we obtain
\begin{align}
\frac{1}{2\ln 2}\left(\frac {\beta_k}{1+\beta_k x_k}-\frac {\beta_k}{1+(1-\varepsilon)\rho_1\alpha_k+\beta_k x_k}\right)
\notag \\
-\nu+\lambda_k=0, \forall k.
\end{align}
Due to the fact that $\lambda_k \ge 0$, it holds that
\begin{align}\label{hx}
\nu \ge \frac{1}{2\ln 2} \frac {\frac{(1-\varepsilon)\rho_1\alpha_k}{\beta_k}}{(x_k+\frac {1}{\beta_k})\left(x_k+\frac {(1-\varepsilon)\rho_1\alpha_k+1}{\beta_k}\right)}, \forall k.
\end{align}

Considering (\ref {R6_1}), we have
\begin{align}
x_k\left[\nu-\frac{1}{2\ln 2} \frac {\frac{(1-\varepsilon)\rho_1\alpha_k}{\beta_k}}{(x_k+\frac {1}{\beta_k})\left(x_k+\frac {(1-\varepsilon)\rho_1\alpha_k+1}{\beta_k}\right)}\right]=0.
\end{align}

Then following the similar steps as in \cite{swipt_2}, the optimal $x_k$ can be derived as
\begin{align}\label{x}
x_k=&\frac{1}{2\beta_k}[\sqrt {(1-\varepsilon)^2\rho_1^2\alpha_k^2+ \frac{2}{\ln 2}(1-\varepsilon)\rho_1\alpha_k\beta_k \mu}\notag \\
& -(1-\varepsilon)\rho_1\alpha_k-2 ]^+
\end{align}
where $(a)^+=\max\{0,a\}$ and $\mu=\frac{1}{\nu}$ can be obtained from (\ref {R6_3}). As such
\begin{align}
l(\mu)=&\frac{\rho_1}{2 \ln 2}\sum_{k=1}^D\left[\left(\frac {1}{1+(1-\varepsilon) \rho_1\alpha_k+\beta_k x_k} \right. \right.\notag\\
&\left. \left. -\frac {1}{1+(1-\varepsilon) \rho_1\alpha_k}\right)\alpha_k\right]+\frac {1}{\mu} \eta \rho_2\alpha_k=0.
\end{align}

Due to the inter-depended relationships among $x_k, \varepsilon$, and $\nu$, it is difficult to derive the optimal closed form expressions for all the variables at the same time. To solve the problem, here we introduce an iterative method by firstly fixing $\varepsilon$ in each iteration.

Now we need to check the availability of root-searching for $l(\mu)=0$. Obviously, $l(\mu)$ descends when $\mu \in \left[\max_k 2\ln 2 \frac {1+(1-\varepsilon)\rho_1\alpha_k}{(1-\varepsilon)\rho_1\alpha_k\beta_k} ,~\infty \right)$. Also, we notice that when $\mu \in \left[\min_k 2\ln 2 \frac {1+(1-\varepsilon)\rho_1\alpha_k}{(1-\varepsilon)\rho_1\alpha_k\beta_k} ,\max_k 2\ln 2 \frac {1+(1-\varepsilon)\rho_1\alpha_k}{(1-\varepsilon)\rho_1\alpha_k\beta_k}\right]$, $l(\mu)$ still decreases since in this interval $x_k $ either remains $0$ or increases with $\mu$. To be exact, we have
\begin{align}
l(\infty) \to -\frac{\rho_1}{2 \ln 2}\sum_{k=1}^D \frac{\alpha_k}{1+(1-\varepsilon)\rho_1\alpha_k} <0,
\end{align}
and
\begin{align}
l&\left(\min_k 2\ln 2\frac {1+(1-\varepsilon)\rho_1\alpha_k}{(1-\varepsilon)\rho_1\alpha_k\beta_k}\right)  \notag\\
&=\max_k \frac{1}{2\ln 2} \frac {(1-\varepsilon)\rho_1\alpha_k\beta_k}{1+(1-\varepsilon)\rho_1\alpha_k}  \sum_{k=1}^D \eta\rho_2\alpha_k >0.
\end{align}
In contrast, it is obvious that $x_k=0, \forall k$, when $\mu \in \left(0, \min_k 2\ln 2 \frac {1+(1-\varepsilon)\rho_1\alpha_k}{(1-\varepsilon)\rho_1\alpha_k\beta_k}\right)$, and thus we know that
\begin{align}
l(u)=&\frac{1}{\mu} \sum_{k=1}^D \eta\rho_2\alpha_k >0.
\end{align}

Consequently, an optimal $\mu^*$ satisfying $l(\mu^*)=0$ can always be found within $\left(\min_k 2\ln 2 \frac {1+(1-\varepsilon)\rho_1\alpha_k}{(1-\varepsilon)\rho_1\alpha_k\beta_k}, \infty \right)$ by root-finding strategies such as bisection searching. Finally, with $x_k, \forall k,$ known, we can calculate the optimal PS ratio using (\ref {R6_o}) as follows
\begin{align}\label{rho}
\varepsilon=\frac{\sum_{k=1}^D {x_k}}{\eta \rho_2 \sum_{k=1}^D \alpha_k}
\end{align}

The iteration framework is summarized in Algorithm~1.
\begin{algorithm}
    \caption{Iteration scheme for uniform source precoding}
  \begin{algorithmic}[1]
	\STATE \textbf{Initialization} let $\varepsilon=0.001$
	\WHILE{$\varepsilon < 1$}
      \STATE  find an optimal $\mu^*$ to make $l(\mu^*)=0$ using root-finding method
	    \STATE  calculate the corresponding $x_k$ using (\ref {x})
			\STATE  calculate the optimal $\varepsilon^*$ using (\ref {rho})
			\STATE  if $|\varepsilon^*-\varepsilon|$ is small enough, iteration terminates. Otherwise, $\varepsilon=\varepsilon+0.001$
    \ENDWHILE
  \end{algorithmic}
\end{algorithm}

\section{Joint Source, Relay and PS ratio optimization}\label{sec_algo_joint}
In this section, we consider a more general scenario of optimal source covariance matrix and thus the source covariance matrix, relay beamforming matrix and PS ratio need to be jointly optimized.

In this case, the achievable rate of the MIMO relay system with PS based energy harvesting can be expressed as
\begin{equation}
C=\frac {1}{2}\log_2 {\rm det} \left({\bf I}_D+(1-\varepsilon) \frac{{\bf H}_2 {\bf F}{\bf H}_1 {\bf Q} {\bf H}_1^H{\bf F}^H{\bf H}_2^H}{\sigma_2^2{\bf I}_D+\sigma_1^2{\bf H}_2 {\bf F}{\bf F}^H{\bf H}_2^H}\right).
\end{equation}

Then the capacity maximization problem with power constraints at both the source and relay nodes can be written as
\begin{subequations}\label{maxRobD}
\begin{align}
\max_{{\bf F},{\bf Q}, \varepsilon}~~& C\\
{\rm s.t.}~~&{\rm tr}{({\bf Q})} \le P,\\
&{\rm tr}{(\sigma_1^2{\bf F}{\bf F}^H +(1-\varepsilon){\bf F}{\bf H}_1 {\bf Q} {\bf H}_1^H{\bf F}^H)} \notag \\
& \qquad \qquad\le \varepsilon \eta {\rm tr}{({\bf H}_1 {\bf Q} {\bf H}_1^H)}. \label{R1_1}
\end{align}
\end{subequations}

To take the advantages of the results already obtained in the uniform source precoding case, we introduce an equivalent channel $\widehat{{\bf H}}_1={\bf H}_1 {\bf Q}^{\frac {1}{2}}$ and thus find that the structure of the optimal relay beamforming still works. That is to say, $\widehat{{\bf F}}={\bf V}_2 \widehat{{\bf \Sigma}}_F \widehat{{\bf U}}_1^H$. Meanwhile, $\widehat{{\bf \Sigma}}_F$ is diagonal, and $\widehat{{\bf U}}_1$ comes from the SVD $\widehat{{\bf H}}_1=\widehat{{\bf U}}_1 \widehat{{\bf \Sigma}}_1 \widehat{{\bf V}}_1^H$. Moreover, because the objective function and the transmit power constraint at relay only depend on $\widehat{{\bf \Sigma}}_1$ but not on $\widehat{{\bf U}}_1$, it was claimed in \cite{swipt_3} that the optimal ${\bf Q}$ must require the least transmit power. Although power splitting based energy harvesting is introduced, it can be proved that the structures of the optimal source covariance and relay beamforming matrices in \cite{swipt_3} still work by defining a new variable $\widehat{\rho}_1=(1-\varepsilon)\rho_1$. So the optimal structures of the source and relay matrices in (\ref{maxRobD}) can be written as
\begin{align}
{\bf F}={\bf V}_2 {\bf \Sigma }_F {\bf U}_1^H,\label{structure1}\\
{\bf Q}={\bf V}_1 {\bf \Lambda }_Q {\bf V}_1^H,\label{structure2}
\end{align}
where ${\bf \Sigma}_F, {\bf \Lambda}_Q$ are diagonal matrices. ${\bf U}_1, {\bf V}_1,{\bf U}_2, {\bf V}_2$ have been defined in (\ref{channel}) and (\ref{channel1}). Then we let ${\bf \Lambda_Q}={\rm diag}({q_1,q_2,...,q_D})$, and ${\bf \Lambda_F}={\bf \Sigma}_F^2={\rm diag}({f_1,f_2,...,f_D})$. Substituting (\ref {structure1}) and (\ref {structure2}) into (\ref {maxRobD}) and introducing a set of new variables $d_k=f_k((1-\varepsilon)\alpha_kq_k+{\bf \sigma}_1^2), \forall k$, we can then rewrite the optimization problem (\ref{maxRobD}) as
\begin{subequations}\label{maxRobe1}
\begin{align}
\max_{\varepsilon,\{d_k\},\{q_k\}} &\tilde{f}(\varepsilon,\{d_k\},\{q_k\})\label{Rd_o}\\
{\rm s.t.}~~&\sum_{k=1}^D  q_k \le P, \label{Rd_1} \\
&\tilde{g} (\varepsilon,\{d_k\},\{q_k\}) \ge 0, \label{Rd_2} \\
&q_k \ge 0, d_k \ge 0, \forall k \\
&0 \le \varepsilon \le 1,
\end{align}
\end{subequations}
where we have defined
\begin{align}
&\tilde{f}(\varepsilon,\{d_k\},\{q_k\}) \triangleq \frac {1}{2}\sum_{k=1}^D\log_2 \frac {\left(1+(1-\varepsilon)\frac{\alpha_k}{\sigma_1^2}q_k\right)\left(1+\frac{\beta_k}{\sigma_2^2}d_k\right)}{1+(1-\varepsilon)\frac{\alpha_k}{\sigma_1^2}q_k+\frac{\beta_k}{\sigma_2^2}d_k},\\
&~~~~~~~~~\tilde{g} (\varepsilon,\{d_k\},\{q_k\})\triangleq \varepsilon \eta\sum_{k=1}^D \alpha_k q_k - \sum_{k=1}^D  d_k.
\end{align}
Note that (\ref{maxRobe1}) involves only scalar variables compared with the matrix variables in (\ref{maxRobD}). However, the problem is still non-convex and it is difficult to obtain a closed-form solution. In the following, we propose an iterative algorithm which can be proved to converge at least to a local optimal solution. For notational simplicity, we let $\boldsymbol{q}=[q_1,q_2,\dots,q_D]^T$, and $\boldsymbol{d}=[d_1, d_2,\dots,d_D]^T$.

\subsection{Optimization with fixed $\boldsymbol{q}$}

We first fix $\boldsymbol{q}$ satisfying (\ref {Rd_1}) and search for the optimal corresponding $\boldsymbol{d}$ and $\varepsilon$. Considering the Lagrangian of (\ref{maxRobe1}), we formulate the following dual problem
\begin{subequations}\label{maxRobe}
\begin{align}
\max_{\varepsilon,\{d_k\},\atop \nu,\{\lambda_k\}} {\cal L} \triangleq \tilde{f}(\varepsilon,\{d_k\})+\nu \tilde{g}(\varepsilon,\{d_k\})+\sum_{k=1}^D  \lambda_k d_k \\
{\rm s.t.}~0 \le \varepsilon \le 1, d_k \ge 0, \nu \ge 0, \lambda_k \ge 0,\forall k.
\end{align}
\end{subequations}
The corresponding KKT conditions can be written as
\begin{subequations}\label{minRobh}
\begin{align}
\nu \tilde{g}(\varepsilon,\{d_k\}) =0,&\label{Rh_0}\\
\lambda_k d_k=0,&\forall k,\label{Rh_1}\\
{\nabla_{d_k}}{\cal L}=0,&\forall k,\label{Rh_2}\\
\nabla_\varepsilon{\cal L}=0,&\label{Rh_3}
\end{align}
\end{subequations}
Following similar approach as in the fixed source covariance matrix case, the optimal $d_k$ can be derived as
\begin{align}\label{expd2}
d_k=&\frac{\sigma_2^2}{2\beta_k}\left(\sqrt{(1-\varepsilon)^2(\frac{\alpha_k}{\sigma_1^2}q_k)^2+2(1-\varepsilon)\frac{\alpha_k}{\sigma_1^2 \ln 2}q_k\beta_k \mu}\right. \notag \\
&\left. -(1-\varepsilon)\frac{\alpha_k}{\sigma_1^2}q_k-2\right)^+,
\end{align}
where $\mu=\frac{1}{\nu}$ can be obtained using (\ref {Rh_3}). Thus we have
\begin{align}\label{expd3}
l(\mu)=&-\frac{1}{2\ln 2}\left[\sum_{k=1}^D \frac{\alpha_k}{\sigma_1^2}q_k\left(\frac{1}{1+(1-\varepsilon)\frac{\alpha_k}{\sigma_1^2}q_k}\right. \right.\notag \\
&\left. \left.-\frac{1}{1+(1-\varepsilon)\frac{\alpha_k}{\sigma_1^2}q_k+\frac{\beta_k}{\sigma_2^2}d_k}\right)\right]+\frac{1}{\mu} \eta \sum_{k=1}^D\alpha_k q_k=0.
\end{align}
Note that in this case, both $\varepsilon$ and $\mu$ are needed to calculate $d_k$. Here we introduce an initial $\varepsilon$, and then search for the optimal $\mu$ by bisection method using (\ref {expd2}) and (\ref {expd3}). With $\mu$ known, we calculate corresponding $d_k$ using (\ref {expd2}) and then the optimal $\varepsilon^*$ based on (\ref {Rh_0}) as follows
\begin{align} \label{rho1}
\varepsilon^* =\frac {\sum_{k=1}^D  {d_k}}{\eta \sum_{k=1}^D  {\alpha_k q_k}}.
\end{align}
The iterative framework is similar to Algorithm~1 in the previous section, and is ignored to avoid redundancy.

\subsection{Optimization with fixed $\boldsymbol{d}$ and $\varepsilon$}

With fixed $\boldsymbol{d}$ and $\varepsilon$, we tackle the more challenging task of calculating $q_k$.
Considering the Lagrangian of problem (\ref {maxRobe1}), the dual problem is defined as
\begin{subequations}\label{maxRobi}
\begin{align}
\max_{\varepsilon,\{q_k\},\nu_1,\atop \nu_2,\{\lambda_k\}} {\cal L}=&\tilde{f}(\varepsilon,\{q_k\})+\nu_1 (P-\sum_{k=1}^D  q_k) \notag \\
&+\nu_2 \tilde{g}(\rho,\{q_k\})+\sum_{k=1}^D  {\lambda_k q_k} \\
{\rm s.t.}~~0 \le \varepsilon \le 1, &\nu_1 \ge 0, \nu_2 \ge 0, \lambda_k \ge 0, q_k \ge 0, \forall k.
\end{align}
\end{subequations}
The corresponding KKT conditions are given by
\begin{subequations}\label{minRobj}
\begin{align}
\nu_1 (P-\sum_{k=1}^D  q_k)=0,&\label{Rj_0}\\
\nu_2 \tilde{g}(\varepsilon, \{q_k\})=0,&\label{Rj_1}\\
\lambda_k q_k=0,& \forall k,\label{Rj_2}\\
{\nabla_{q_k}}{\cal L}=0,&\forall k,\label{Rj_3}\\
\nabla_\varepsilon{\cal L}=0.&\label{Rj_4}
\end{align}
\end{subequations}
Then according to (\ref {Rj_3}), we have
\begin{align}
\frac{(1-\varepsilon)\alpha_k}{2\ln 2~\sigma_1^2}&\left(\frac{1}{1+(1-\varepsilon)\frac{\alpha_k}{\sigma_1^2}q_k}-\frac{1}{1+(1-\varepsilon)\frac{\alpha_k}{\sigma_1^2}q_k+\frac{\beta_k}{\sigma_2^2}d_k}\right) \notag \\
&~~~~~~-\nu_1+\nu_2\varepsilon\eta\alpha_k+\lambda_k=0.
\end{align}
Let $\hat{\nu}_k=2(\nu_1-\nu_2\varepsilon\eta\alpha_k)$. From (\ref {Rj_3}) we then have
\begin{align}\label{expq2}
q_k=&\frac{\sigma_1^2}{2\alpha_k(1-\varepsilon)}\left(\sqrt {(\frac{\beta_k}{\sigma_2^2}d_k)^2+4 \frac{(1-\varepsilon)\alpha_k\beta_k}{\ln 2 \sigma_1^2\sigma_2^2}d_k\hat{\mu}_k} \right. \notag \\
&\left. -\frac{\beta_k}{\sigma_2^2}d_k-2\right)^+,
\end{align}
where $\hat{\mu}_k=\frac{1}{\hat{\nu}_k}$.

Since each $q_k$ depends on its own dual variable $\hat{\mu}_k$, it is difficult to find all the dual variables by searching. Instead, here we use the dual decomposition method proposed in \cite{swipt_12} to find the optimal solution to the dual problem (\ref {maxRobi}). The key idea is to find the optimal dual variables $\nu_1$ and $\nu_2$ by searching in turn until finding out the possible values that satisfy both constraints. Then we use them to calculate the corresponding $q_k$. The detailed algorithm is formally presented in Algorithm~2.

\begin{algorithm} 
    \caption{Dual Decomposition for PS Relaying}
  \begin{algorithmic}[1]
    \STATE \textbf{Main Function}
		\STATE  Fix $\boldsymbol{d}$ and $\varepsilon$
		\STATE  $\boldsymbol{q}=optimize$~$\nu_1(\boldsymbol{d}, \varepsilon)$
    \STATE  \textbf{Function} $\boldsymbol{q}=optimize$~$\nu_1(\boldsymbol{d}, \varepsilon)$
		\STATE  $\phi_k=\frac{(1-\varepsilon) \alpha_k \beta_k d_k}{2\sigma_1^2\ln 2(\sigma_2^2+\beta_k d_k)},~\nu_{1min}=\nu_{1max}=\max_k(\boldsymbol{\phi})$
    \WHILE{$\sum_{k=1}^D q_k \ge P$}
      \STATE $\nu_{1max}=\nu_{1max}+10^{-4}$
	    \STATE $q=optimize~\nu_2(\nu_{1max}, \boldsymbol{d}, \varepsilon)$
    \ENDWHILE
		 \WHILE{$|\nu_{1max}-\nu_{1min}|>\varepsilon$}
      \STATE  $\nu_1=\frac{\nu_{1max}+\nu_{1min}}{2}$
	    \STATE  $q=optimize~\nu_2(\nu_1,\boldsymbol{d}, \varepsilon)$
			\STATE  if $\sum_{k=1}^D q_k \ge P$, $\nu_{1min}=\nu_1$; otherwise, $\nu_{1max}=\nu_1$
    \ENDWHILE
		\STATE \textbf{Function} $\boldsymbol{q}=optimize~\nu_2(\nu_1,\boldsymbol{d}, \varepsilon)$
		\STATE  $\theta_k=\frac{\nu_1-\phi_k}{\eta \varepsilon \alpha_k},~\nu_{2min}=\nu_{2max}=\min_k(\boldsymbol{\theta})$	
    \WHILE{$ \eta \varepsilon \sum_{k=1}^D \alpha_k q_k-\sum_{k=1}^D d_k \le 0$}
      \STATE $\nu_{2max}=\nu_{2max}+10^{-4}$
	    \STATE $\boldsymbol{q}=optimize(\nu_1, \nu_{2max},\boldsymbol{d}, \varepsilon)$
    \ENDWHILE
		 \WHILE{$|\nu_{1max}-\nu_{1min}|>\varepsilon$}
      \STATE  $\nu_1=\frac{\nu_{1max}+\nu_{1min}}{2}$
	    \STATE  $\boldsymbol{q}=optimize~\nu_2(\nu_1,\boldsymbol{d}, \varepsilon)$
			\STATE  if $ \eta \varepsilon \sum_{k=1}^D  \alpha_k q_k-\sum_{k=1}^D  d_k \le 0$, let $\nu_{2min}=\nu_2$; otherwise, $\nu_{2max}=\nu_2$
    \ENDWHILE
		\STATE \textbf{Function} $\boldsymbol{q}=optimize (\nu_1,\nu_2,\boldsymbol{d}, \varepsilon)$
		\STATE calculate $\boldsymbol{q}$ according to (\ref {expq2})
  \end{algorithmic}
\end{algorithm}
	
\subsection{Iterative Optimization}
Algorithm~3 presents the framework of iteration to solve problem (\ref{maxRobe1}).

\begin{algorithm} 
    \caption{Iteration Framework for PS Relaying}
  \begin{algorithmic}[1]
	  \STATE \textbf{Initialization} Let $\boldsymbol{q}$ satisfying (\ref {Rd_1})
		\STATE  Calculate optimal $\boldsymbol{d}$ and $\varepsilon$ with fixed $\boldsymbol{q}$ using (\ref {expd2}) and Algorithm 1
		\STATE  Re-optimize $\boldsymbol{q}$ with the obtained $\boldsymbol{d}$ and $\varepsilon$ via dual decomposition method in \textbf{Algorithm~2}
    \STATE  Return to Step 2 until convergence
  \end{algorithmic}
\end{algorithm}

\section{Numerical Results}\label{sec_sim}
In this section, simulation is done to analyse the performance of the proposed approaches for an energy harvesting enabled MIMO relay system. The achievable capacity and optimal PS ratio are investigated against different values of the noise variances, $\sigma_1^2$ at relay and $\sigma_2^2$ at destination. Here we use Rician fading channels for channel modeling. Both ${\bf H_1}$ and ${\bf H_2}$ are modeled with a set of independent zero-mean complex Gaussian random variables with a variance of $20{\rm dBm}$. The maximum source transmit power $P$ is set to be $30{\rm dBm}$. Unless otherwise stated, the terminating threshold for iteration is $10^{-3}$. Case I denotes the fixed source covariance matrix scenario with uniform source precoding while Case II presents the joint source, relay and PS ratio optimization.

\begin{figure}
\centering
\includegraphics*[width=6cm]{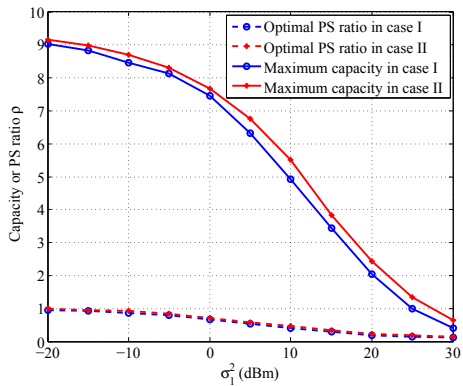}
\caption{Optimal PS ratio and capacity versus $\sigma_1^2$ with $\sigma_2^2=0{\rm dBm}$.}
\label{simu1}
\end{figure}

Fig.~\ref{simu1} presents how the value of the noise variance $\sigma_1^2$ at relay decides the maximum capacity and optimal PS ratio for the proposed iterative schemes. The noise variance $\sigma_1^2$ at relay varies from $-20{\rm dBm}$ to $30{\rm dBm}$ while the noise variance $\sigma_2^2$ at destination is fixed at $\sigma_2^2=0{\rm dBm}$. As can be observed, the joint source, relay and PS ratio optimization shows capacity gain over the uniform source precoding case. The optimal PS ratio in Case II is also a little higher. And in both cases, the maximum capacity and optimal PS ratio decrease dramatically with the increase of the noise variance $\sigma_1^2$ at the relay node.

\begin{figure}
\centering
\includegraphics*[width=6cm]{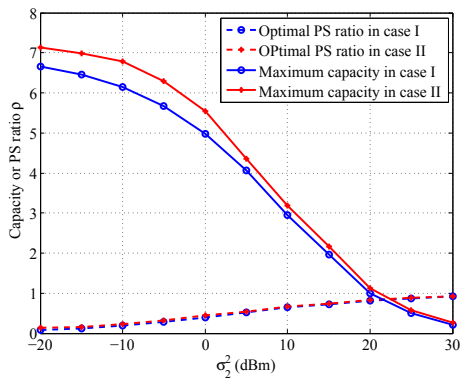}
\caption{Optimal PS ratio and capacity versus $\sigma_2^2$ with $\sigma_1^2=10{\rm dBm}$.}
\label{simu2}
\end{figure}

Fig.~\ref{simu2} shows the maximum capacity and optimal PS ratio for the proposed schemes versus the value of the noise variance $\sigma_2^2$ at destination. Here we let the noise variance $\sigma_2^2$ vary from $-20{\rm dBm}$ to $30{\rm dBm}$ with the noise variance $\sigma_1^2$ fixed at $10{\rm dBm}$. According to the figure, Case II always outperforms the uniform source precoding case with an extremely obvious capacity gain at a low noise variance $\sigma_2^2$, e.g. $-20{\rm dBm}$. Also, the maximum capacity drops with increasing noise variance $\sigma_2^2$ in both cases. On the contrary, the optimal PS ratio rises with the increase of the noise variance $\sigma_2^2$ from $-20{\rm dBm}$ to $30{\rm dBm}$ which is different from what we see in Fig.~\ref{simu1}.

\section{Conclusion}\label{sec_con}
This paper investigates wireless information and power transfer for MIMO power splitting relaying. Aiming to maximize the capacity subject to power constraints at the source and relay, joint optimization of the source, relay and PS ratio as well as the fixed source covariance matrix scenario are considered. Instead of semidefinite relaxation (SDR) based solutions which rely on existing software solvers, here we introduce the structures of the optimal source covariance and relay precoding matrices and then turn the matrix optimization problem into scalar optimization. Iterative schemes are proposed to yield near-optimal solutions in both cases. Finally, the performance of the proposed schemes are investigated via simulations.

\bibliographystyle{IEEEtran}\footnotesize


\end{document}